\documentstyle[times,11pt]{article}
\begin{document}
\textwidth 12.7cm
\textheight 20.4cm
\headsep 6mm
\topskip 11pt
\leftmargin 0.5cm
\flushbottom
\mathsurround 1pt

\renewcommand{\textfraction}{0}
\renewcommand{\topfraction}{1}
\renewcommand{\floatpagefraction}{0.5}
\setcounter{secnumdepth}{3}

\newcommand{\dd}{\,{\rm d}}
\newcommand{\ie}{{\it i.e.},\,}
\newcommand{\etal}{{\it et al.\ }}
\newcommand{\eg}{{\it e.g.},\,}
\newcommand{\cf}{{\it cf.\ }}
\newcommand{\vs}{{\it vs.\ }}
\newcommand{\ttt}[1]{\!\times\!10^{#1}}
\newcommand{\sst}{\scriptscriptstyle}
\newcommand{\fns}{\footnotesize}
\newcommand{\beq}{\begin{equation}
  \renewcommand{\int}{\intop\limits}
  \renewcommand{\oint}{\ointop\limits}}
\newcommand{\eeq}{\end{equation}}
\newcommand{\beqarr}{\par\begin{minipage}{11cm} \begin{eqnarray*}}
\newcommand{\eeqarr}{\end{eqnarray*} \end{minipage} \hfill 
   \stepcounter{equation}{\rm (\theequation)}\vspace{3mm}\linebreak}
\newcommand{\bdm}{\begin{displaymath}
  \renewcommand{\int}{\intop\limits}
  \renewcommand{\oint}{\ointop\limits}}
\newcommand{\edm}{\end{displaymath}}
\newcommand{\trule}{\rule{0pt}{14pt}}
\newcommand{\zdot}{\makebox[0pt][l]{.}}
\newcommand{\up}[1]{\ifmmode^{\rm #1}\else$^{\rm #1}$\fi}
\newcommand{\dn}[1]{\ifmmode_{\rm #1}\else$_{\rm #1}$\fi}
\newcommand{\upd}{\up{d}}
\newcommand{\uph}{\up{h}}
\newcommand{\upm}{\up{m}}
\newcommand{\ups}{\up{s}}
\newcommand{\arcd}{\ifmmode^{\circ}\else$^{\circ}$\fi}
\newcommand{\arcm}{\ifmmode{'}\else$'$\fi}
\newcommand{\arcs}{\ifmmode{''}\else$''$\fi}
\newcommand{\pder}[2]{\frac{\partial#1}{\partial#2}}
\newcommand{\der}[2]{\frac{{\rm d}#1}{{\rm d}#2}}
\newcommand{\m}{{\rm m}}
\newcommand{\cm}{{\rm cm}}
\newcommand{\g}{{\rm g}}
\newcommand{\kg}{{\rm kg}}
\newcommand{\s}{{\rm s}}
\newcommand{\MS}{{\rm M}\ifmmode_{\odot}\else$_{\odot}$\fi}
\newcommand{\RS}{{\rm R}\ifmmode_{\odot}\else$_{\odot}$\fi}
\newcommand{\LS}{{\rm L}\ifmmode_{\odot}\else$_{\odot}$\fi}
\newcounter{pagefrom}
\newcounter{pageto}
\newcounter{volume}
\newcounter{year}

\newcommand{\SetPages}[2]{\setcounter{page}{#1}
  \setcounter{pagefrom}{#1}\setcounter{pageto}{#2}}
\newcommand{\SetVol}[2]{\setcounter{volume}{#1}\setcounter{year}{#2}}
\newcommand{\SetNum}[1]{\newcommand{\Num}{{\bf #1}}}

\newenvironment{Titlepage}{
\vspace*{2cm}
  \begin{center}
}{
  \end{center}\par\vspace{3mm}
}

\newcommand{\Title}[1]{{\large\bf\boldmath #1 \\[3mm] {\footnotesize by} 
\\[3mm]}}

\newcommand{\Author}[2]{{\large\spaceskip 2pt plus 1pt minus 1pt #1}\\[3mm]
   {\small #2}\\[6mm]}

\newcommand{\Received}[1]{}

\newcommand{\Abstract}[2]{{\footnotesize\begin{center}ABSTRACT\end{center}
\vspace{1mm}\par#1\par
\noindent
{\bf Key words:~~}{\it #2}}}

\newcommand{\FigCap}[1]{\footnotesize\par\noindent Fig.\  %
  \refstepcounter{figure}\thefigure. #1\par}

\newcommand{\TabCap}[2]{\begin{center}\parbox[t]{#1}{\begin{center}
  \small {\spaceskip 2pt plus 1pt minus 1pt T a b l e}
  \refstepcounter{table}\thetable \\[2mm]
  \footnotesize #2 \end{center}}\end{center}}

\newcommand{\Table}[3]{\begin{table}[htb]\TabCap{#2}{#3}
  \vspace{#1}\end{table}}

\newcommand{\TableSep}[2]{\begin{table}[p]\vspace{#1}
\TabCap{#2}\end{table}}

\newcommand{\TableFont}{\footnotesize}
\newcommand{\TableFontIt}{\ttit}
\newcommand{\SetTableFont}[1]{\renewcommand{\TableFont}{#1}}

\newcommand{\MakeTable}[4]{\begin{table}[htb]\TabCap{#2}{#3}
  \begin{center} \TableFont \begin{tabular}{#1} #4 
  \end{tabular}\end{center}\end{table}}

\newcommand{\MakeFrameTable}[5]{\begin{table}[htb]\TabCap{#2}{#3}
  \begin{center} \TableFont \begin{tabular}{#1}\hline\trule #4\\[1mm]
  \hline\trule #5\\[1mm]\hline
  \end{tabular}\end{center}\end{table}}

\newcommand{\MakeOwnTable}[6]{\begin{table}[htb]\TabCap{#2}{#3}
  \begin{center}\TableFont #4 \begin{tabular}{#1} #5
  \end{tabular} #6 \end{center}\end{table}}

\newcommand{\MakeTableSep}[4]{\begin{table}[p]\TabCap{#2}{#3}
  \begin{center} \TableFont \begin{tabular}{#1} #4 
  \end{tabular}\end{center}\end{table}}

\newcommand{\Figure}[2]{\begin{figure}[htb]\vspace{#1}
\FigCap{#2}\end{figure}}

\newcommand{\FigureSep}[2]{\begin{figure}[p]\vspace{#1}
\FigCap{#2}\end{figure}}

\newcommand{\FigureOwn}[3]{\begin{figure}[htb]\vspace{#1}#3
\FigCap{#2}\end{figure}}
\newcommand{\FigureOwnSep}[3]{\begin{figure}[p]\vspace{#1}#3
\FigCap{#2}\end{figure}}
\newcommand{\LeavePage}{\begin{table}[p]\vspace{19.5cm}\end{table}}

\newenvironment{references}%
{
\footnotesize \frenchspacing
\renewcommand{\thesection}{}
\renewcommand{\in}{{\rm in }}
\renewcommand{\AA}{Astron.\ Astrophys.}
\newcommand{\AAS}{Astron.~Astrophys.~Suppl.~Ser.}
\newcommand{\ApJ}{Astrophys.\ J.}
\newcommand{\ApJS}{Astrophys.\ J.~Suppl.~Ser.}
\newcommand{\ApJL}{Astrophys.\ J.~Letters}
\newcommand{\AJ}{Astron.\ J.}
\newcommand{\IBVS}{IBVS}
\newcommand{\PASP}{P.A.S.P.}
\newcommand{\Acta}{Acta Astron.}
\newcommand{\MNRAS}{MNRAS}
\renewcommand{\and}{{\rm and }}
\section{{\rm REFERENCES}}
\sloppy \hyphenpenalty10000
\begin{list}{}{\leftmargin1cm\listparindent-1cm
\itemindent\listparindent\parsep0pt\itemsep0pt}}%
{\end{list}\vspace{2mm}}

\def\TYLDA{~}
\newlength{\DW}
\settowidth{\DW}{0}
\newcommand{\dw}{\hspace{\DW}}

\newcommand{\refitem}[5]{\item[]{#1} #2%
\def\REFARG{#3}\ifx\REFARG\TYLDA\else, {\it#3}\fi
\def\REFARG{#4}\ifx\REFARG\TYLDA\else, {\bf#4}\fi
\def\REFARG{#5}\ifx\REFARG\TYLDA\else, {#5}\fi.}

\newcommand{\Section}[1]{\section{\normalsize\bf#1}}
\newcommand{\Subsection}[1]{\vspace*{-1mm}\subsection{\normalsize\em{#1}}}
\newcommand{\Acknow}[1]{\par\vspace{5mm}{\bf Acknowledgements.} #1}
\pagestyle{myheadings}

\newfont{\bb}{timesbi at 12pt}
\newcommand{\coo}[7]{$\alpha_{#1}=#2\up{h}#3\up{m}#4\up{s}$, 
$\delta_{#1}=#5\arcd#6\arcm#7\arcs$}
\begin{Titlepage}
\Title{The All Sky Automated Survey\footnote{Based on observations obtained 
at the Las Campanas Observatory of the Carnegie Institution of Washington.}}

\Author{G.~~P~o~j~m~a~\'n~s~k~i}{Warsaw University Observatory
Al~Ujazdowskie~4, 00-478~Warsaw, Poland}

\Received{September 30, 1990}

\end{Titlepage}

\Abstract{Technical description of the new project called All Sky Automated 
Survey and results of the tests of our prototype instrument are presented. The 
ultimate goal of this project is photometric monitoring of the large area of 
the sky with fully automated, low cost instruments. Possible applications of 
the project are indicated and future prospects discussed. At present over 
hundred square degrees is observed 5--10 times each night in $I$ band, allowing 
us to monitor over 30~000 stars brighter than 12--13~mag. Full description, 
pictures  and current status of the project can be found on the {\em WWW: 
http://sirius.astrouw.edu.pl/$\sim$gp/asas/asas.html}}{Surveys-Telescopes-
Techniques:photometric} 
 
\section{Introduction}
Availability of the large CCD detectors allowed astronomers to start the new 
age of massive survey projects. Most of them propose to search for some kind of 
specific objects or events. Among them are microlensing groups: OGLE (Udalski 
\etal 1992), MACHO (Alcock \etal 1993) and  EROS (Aubourg \etal 1993), 
searches for the optical counterparts of gamma-ray bursts, supernovae searches 
(BAIT, Van Dyk \etal 1994) and projects interested in variable stars, 
near-Earth asteroids, comets, like All Sky Patrol Astrophysics (Braeuer and 
Vogt 1995). At least part of these activities are expected to be performed in 
the automated way. Some are already done by robotic telescopes (Hayes and 
Genet 1989, Schaefer \etal 1994, Henry and Eaton 1995). 

Many interesting scientific programs can be done even with surveys covering 
only the brightest of the observable objects (Paczy{\'n}ski 1997). The main 
topics include variable stars, comets, asteroids and near-Earth objects, 
supernovae or AGNs. 

In this paper we present a description of the new observational project called 
All Sky Automated Survey which we have started at the beginning of 1996. 

The ultimate goal of our project is to detect and investigate any kind of the 
photometric variability present on as much of the sky as possible. At the 
beginning we restrict ourselves only to the brightest few million stars on the 
sky, \ie our magnitude limit is about 13~mag. We will increase this limit in 
future, as our data acquisition and data processing capabilities grow. We want 
to achieve this aim at relatively low cost, using simple  but powerful 
automatic modules linked together at the observing sites with good weather 
conditions. 

\section{Prototype Instrument}
The prototype instrument was built in 1996 at the Warsaw University 
Observatory. It consists of five basic elements: equatorial mount, CCD camera, 
telephoto lens, electronic box and control computer. 

\subsection{Mount}
Equatorial mount is a compact, lightweight (aluminum alloy) horseshoe design 
similar to those built by AutoScope (Genet and Hayes 1989). It consists of the 
base structure with right ascension drive, horseshoe with declination drive 
and camera platform. 

The triangular base rests on the adjustment table. Polar vertex has a precise 
vertical screw used for adjustment of the inclination of polar axis. Spherical 
bearing supporting polar axle is mounted on the upper side of this vertex. The 
other two vertices have horizontal screws used for precise horizontal 
alignment. Two roller holders  extend from these vertices. Each contains two 
bearings and steel roller on which horseshoe rests. One roller runs idle while 
the other, linked to the the RA stepper motor through intermediary sprocket 
gears drives the horseshoe {\it via} friction. 

The horseshoe assembly has three arms of approximately 500~mm length, 
connected at one end to form a polar axle. Two declination axle supports are 
mounted on the front side of the horseshoe while on the back side declination 
sprocket gears and stepper motor are located. 

Stepper motors are driven by the electronic box, containing power supplies, 
microprocessor controllers and translator drives. Both stepper motors are 
controlled over the same serial link. 

RA drive transmission rate (including friction roller and horseshoe) is 
approximately 950:1 which leads to the RA resolution of 3.4 arcsec per step. 
It requires about four steps per second to track the sky rotation. Sprocket 
gears and friction roller form a non-backlash drive with positioning 
repeatability much better than one arcsec. 

Declination drive has much smaller gear rate -- 28.8:1 -- and therefore each 
motor step corresponds to 112.5 arcsec. Positioning accuracy in 
declination at different hour angles is reduced to a few tens of arcsec 
due to the lack of friction element and long elastic chain belts. 

Maximum slewing rate is about 5~deg/sec in hour angle and 150~deg/sec in 
declination. Therefore the average pointing time during a regular observing 
run is about 10 seconds. 

Instrument platform of 280~mm diameter, hanging on two declination 
half-shafts, has mounting holes for three CCD cameras. The prototype has only 
one CCD system mounted. The total weight of the instruments mounted on the 
platform should not exceed 15~kg and the outline of the back side of the 
instrument should stay within the half-sphere of 170~mm radius. 

\subsection{CCD Camera System}
CCD camera is a commercial Meade Pictor 416 product with Kodak KAF~0400 chip. 
It has ${516\times768}$~pixels of 9~micron size with full well capacity of 
85000 electrons. The only available gain of the system is 1.2~$ADU/e^{-}$ and 
therefore pixel counts are easily stored as 2~byte integers. Factory data 
claim 42\% peak quantum efficiency. CCD head has a two-stage thermoelectric 
cooler capable to cool CCD down to 45 degrees below the case temperature. 

Pictor 416 camera has its own electronic box controlling cooling, exposing and 
data acquisition. It is accessible from the interactive PC software {\it via} 
SCSI or RS-232 interfaces. Since we needed a direct control over the camera we 
have added our own driver for the fast (110~kbauds) serial link. This serial 
link is presently a bottle-neck of our system. Transfer time of the full image 
to the main computer is equal to 40-80~sec, while the Pictor electronics 
itself reads CCD chip in 4~seconds only. 

Camera head has built-in two-leaf fast shutter which failed after a few 
thousand exposures. 

We have equipped our camera with 135/1.8 Sigmatel lens which gives the scale 
of 14.2 arcsec/pixel and $3\times2$ deg field of view. Telephoto lens proved 
to have very unpleasant, triangular point spread function but produces very 
little of field distortions. With such fast lens our CCD system is sky limited 
and exposure times are limited to a few minutes. 

There is some space for a filter between the lens and camera head. Currently 
$I$-band filter (Schott RG-9, 3~mm) is used. 

The mount, camera and electronic box are housed in the ${\rm 0.6~m\times0.6~m 
\times0.8~m}$ water-proof ply-wood box, which has to be manually closed in 
case of bad weather. Polar axis of the mount can be adjusted for any latitude 
between 18 and 54 degrees (North or South). For other latitudes special 
supporting frame is needed. 

\subsection{Control Computer}
The control computer is SparcStation 5 with 64~MB RAM, 9~GB disk space and 
DAT-2 tape drive. 

The heart of the control software is a remotely accessible database server, 
which keeps all necessary information about the system and distributes it to 
the interested clients. There are three main database clients: camera driver 
communicating with the Pictor system over the fast serial link, motor driver 
communicating with the mount over the slow serial link and "observer" program 
managing observations. There is also image server sending images to the 
interested clients (usually in the interactive mode) and "analyst" client 
which manages data reduction process. 

\section{Observation Schedule}
Automated observations begin at 5~p.m. when computer initiates "observer" 
program. This in turn performs system checkup and schedules start of 
observation for a few minutes before dusk. Camera cooling is then switched on 
and the system waits for the Sun to hide a few degrees below horizon. At this 
time camera is pointed into zenith and a set of sky flat field images is 
obtained. When the sky is too dark for flat field images, "observer" takes a 
few dark frames and waits until the end of twilight. 

Then regular observations start. "Observer" loops over a list of selected 
fields checking each time visibility conditions. If the field is invisible or 
too low, is obscured by the surrounding obstacles (dome and house) or is too 
close to the Moon -- it is skipped. If visibility conditions are satisfied 
camera is moved into desired position and the shutter is opened. 

Exposure lasts for three minutes. Then camera electronics reads out new frame 
and transfers it to the computer. New image is immediately stored on the disk 
and database server gets informed about completing the observation. This 
information is forwarded to the "analyst" program, which starts frame 
reduction process. Afterwards "observer" checks the Sun position and if it is 
still well below horizon -- goes to the next field on the list. Otherwise it 
starts shutdown procedure and starts data dump to the DAT-2 tape. Such a 
schedule results in about 20 calibration frames and 120 program frames each 
night, creating data stream of 120~MB per night. 

Observing schedule is flexible. In principle it is possible to change or add 
"on fly" observed targets remotely, using Internet and database server. At 
present system requires human interaction only for emergency closing in the 
bad weather conditions and for changing the DAT tape once a month. 

\section{Data Reduction}
At the beginning of the night the "analyst" program waits for the dark 
exposures to be completed. Then it calculates medianed dark frame and, using 
recent sky-flat frames, it computes new flat-field image. Dark and flat-field 
frames are given unique names and stored in a special directory. On some 
nights (\eg if observations start late) new dark or flat frames may not be 
created. 

During regular observations "analyst" sleeps until new frame is saved to the 
disk. Then it performs (depending on the current settings) some of the 
following tasks: dark subtraction, flat-fielding, object detection, 
photometry, astrometry and catalog updating. 

Dark subtraction and flat division are performed using the latest available 
dark and flat images (usually from the beginning of the night). 

Object detection program looks at the image intensity enhancements and tries 
to classify them as stars, galaxies, cosmic rays and various trails. It is 
based on the DAOPHOT algorithm (Stetson 1987), \ie it convolves image with 
the lowered, truncated point spread function and then searches for the local 
maxima above preselected threshold. Object classification is done using 
"sharpness" and "roundness" attributes. Special treatment is required for the 
elongated trails. 

Objects classified as stars and galaxies are then subject to photometry. After 
many experiments with various photometry schemes we decided to stay with 
aperture photometry during the prototype phase of our project. It is much 
faster and more reliable with our non-symmetric PSF (see Section~7.1). 

Astrometry program uses Guide Star Catalog as a source of accurate stellar 
positions. Depending on the field crowding there are between 100 and 1000 GSC 
stars present on each frame. For astrometry solution, which in our case proved 
to require only rotation, translation and scaling, we use only a few hundred 
brightest stars. The rms distance between GSC stars and matching stars usually 
converges to 0.2 pixel (3~arcsec). 

\section{The Catalog}
\subsection{Structure}
Results of the photometry obtained for stars and galaxies are put into the 
photometric catalog which is arranged as a relational database that consists 
of four types of files. 

The first one (FLOG) is an observation log updated as soon as frame exposure 
is completed. Each record (about 100~bytes long) contains sequential frame 
number, time of observation, coordinates of the frame center, camera and site 
information, exposure details, reduction status, quality classification, 
photometry and astrometry transformation coefficients and link to the 
photometry file. 

Main catalog file (CAT) forms a tree structured database in which records (32 
bytes long) are sorted by declination and right ascension. Such construction 
is very flexible although not very fast. New objects are entered at the end of 
the file and only two internal links have to be updated to properly sort the 
database. To access selected object one needs to follow (read) about 
$\log_{2}N$ records from the database (where $N$ is the number of objects in 
the database). At present we are able to search a few hundred objects per 
second. Each CAT record contains $\alpha$ and $\delta$ (2000) coordinates of 
the object, average magnitude and its dispersion, classification and pointer 
to the LINK file. 

Link database (LINK) chains photometry entries for each catalog object. Each 
record (12 bytes long) contains pointer to the FLOG file (giving access to the 
frame details, \eg HJD of exposure and transformation coefficients), pointer 
to the PHOTO file (raw magnitudes) and pointer to the next LINK record for the 
same object. 

Photometry file (PHOTO) is a continuous storage for the photometry results. 
Each record (10 bytes long) contains the following information for the 
detected objects: pointer to the main catalog (CAT) entry, coordinate offsets 
from that entry, raw magnitude and its error, classification details. 

LOG and CAT file are relatively small and therefore only one instant of each 
exists. Photometry and link files grow very fast (a few MB per night) and 
therefore we are prepared to keep multiple PHOTO and LINK files distributed 
over the network. 

\subsection{Catalog Interface and Data Calibration}
Extended C-language library forms an interface to the database which allows 
user to add, delete, search and retrieve data. Many different filters could be 
applied (\eg coordinate and magnitude limits, data scattering, number of 
observations) to retrieve only the most relevant information. Graphical user 
interface on the WWW page can be used to access some catalog data from the 
Internet\\
(http://www.astrouw.edu.pl/$\sim$gp/asas/catalogue.html). 

The most important procedure in the library is "add\_ast" program that puts 
astrometry results into the catalog. Since we want our catalog magnitudes 
$I_{\rm cat}$ to be as close to the standard system as possible, we have to 
convert instrumental magnitudes $i$ using standard transformation (without 
color terms): 
$$I_{\rm cat}=i+i_{0}+k_{I}X\eqno(1)$$
where $i_{0}$ denotes instrumental offset and $k_{I}$ extinction coefficient.

In practice it is enough to perform this calibration only once for each field. 
It requires frames obtained during dark, photometric nights, for which 
extinction coefficient $k_{I}$ is stable and well known. This task has not yet 
been fully automated - requires user to select the best frames for 
calibration. 

Adding photometry results for the already calibrated fields is performed in 
two steps. 

First the catalog is searched for each object on the astrometry list. If 
coordinate match within specified radius (usually 15~arcsec) is found, 
the catalog data are retrieved and tested. If the object contains saturated 
pixels, is unusually elongated, marked as variable or too faint -- it is 
rejected. Otherwise it is added to the matching list. 

In the second step differences ${I_{\rm cat}-i}$ (catalog magnitude {\em 
minus} raw magnitude) are calculated for the matching stars, and the least 
squares method is used to find best $i_{0}$ and $k_{I}$ coefficients. Fit 
residua are then inspected and deviating values rejected. The whole procedure 
is repeated until satisfactory convergence is achieved. 

Such general attitude is required only if substantial differential extinction 
effects are expected. If this is not the case than two other program options 
could be used. First option uses fixed extinction coefficient $k_{I}$ 
(determined from the observation of the same field at different zenith 
distances) and solves transformation equation for $i_{0}$. The second option 
neglects differential extinction on the frame and solves transformation for 
the whole ${\Delta i=i_{0}+k_{I}\overline{X}}$ term, where $\overline{X}$ is 
an average air mass for the frame. 

Field of view of our prototype instrument is ${2\times3}$ degrees and 
therefore differential extinction in $I$-band is always smaller than 0.01~mag. 
Having other sources of larger errors we decided to use last option for 
magnitude transformation. General method will be applied in the next phase of 
experiment, when 2k by 2k CCD detectors will be used. 

Calibration has not to be done immediately after observation. In fact raw 
magnitudes are stored in the PHOT database and only the calibration 
coefficients are saved for each frame in the FLOG records. 

\section{System Integrity Tests}
The prototype was assembled and tested at the Warsaw University Observatory in 
fall/winter 1996. 

CCD tests proved that Pictor 416 system produced acceptable images with 
readout noise of 9 electrons and good linearity. Chip surface was clean and 
only a few dark spots were visible on the flat field images. Electronic 
interference pattern was visible on some dark frames, but its intensity was 
smaller than the read-out noise. Dark frames revealed also some distinct 
populations of hot pixels. At CCD temperature of ${-12\arcd}$C, 6\% of all 
pixels have dark current less than 25~e\up{-}/min, and only 0.1\% larger than 
40~e\up{-}/min. There are also more than 50 pixels with dark current larger 
than 1000~e\up{-}/min The shutter error was not important for exposure times 
longer than 0.5~sec. 

At the beginning of the test the thermocooler was able to cool the CCD 
42{\arcd}C below the case temperature. However, at 80\% cooling efficiency the 
camera head went 10{\arcd}C above the ambient. Therefore small fan was added 
to keep the head case temperature within ambient ${+5}$~deg limit. At the 
average night conditions (${+15\arcd}$C) the CCD detector was kept typically 
at ${-20\arcd}$C. 

\begin{figure}[htb]
\vspace*{5.8cm}
\includegraphics{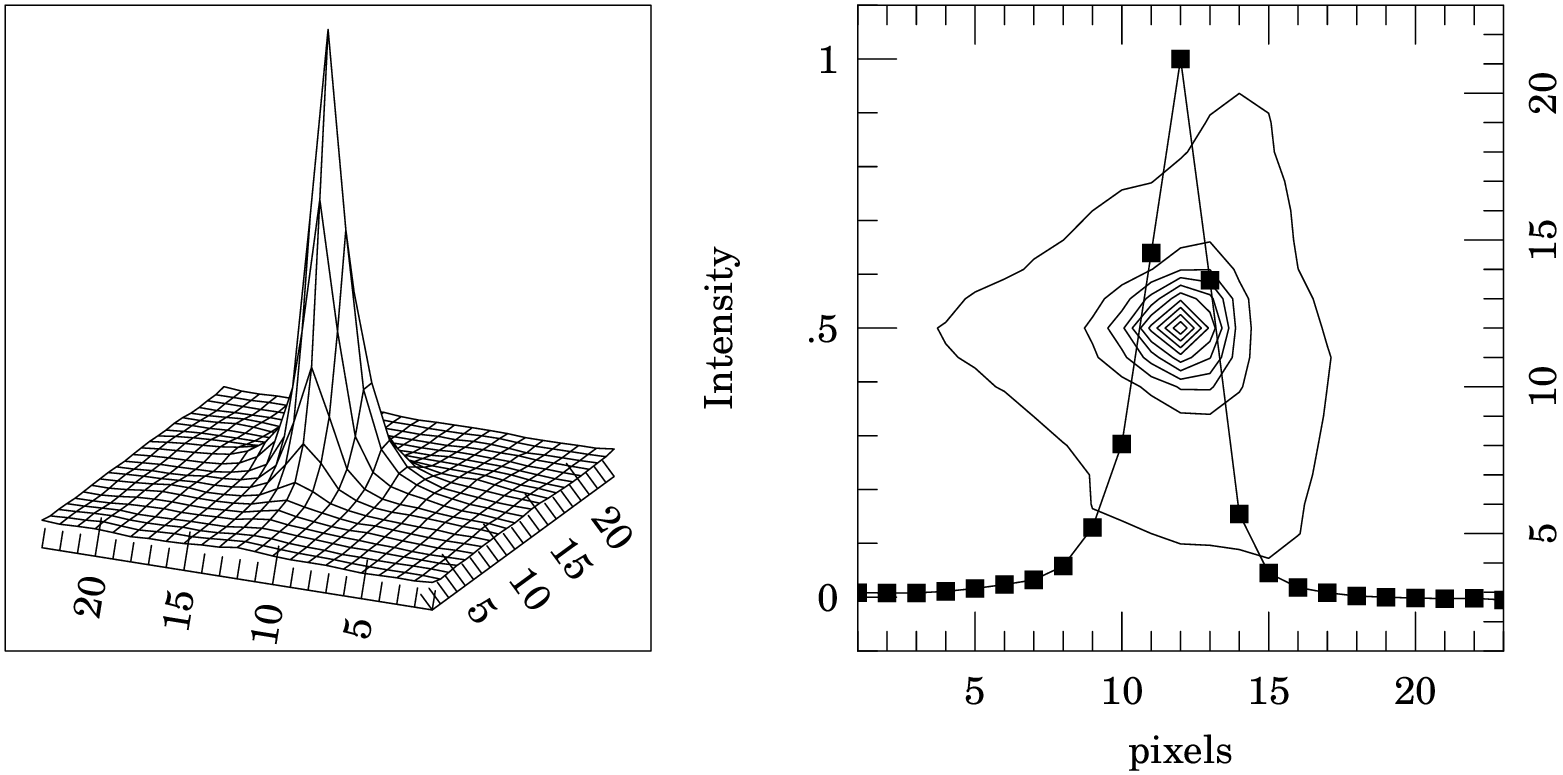}
\FigCap{Point Spread Function of the prototype instrument. Left picture shows 
mesh plot of the PSF (units are 14.2 arcsec pixels). Right picture shows 
contour plot of the triangular PSF (isophotes were plotted from intensity 0.01 
to 1.01 with 0.1 interval) and its intensity cut (black squares).} 
\end{figure}

Due to the huge pixel size seeing is not a limiting factor for our system. At 
14~arcsec per pixel the basic determinant of the PSF size is the lens and 
mount tracking accuracy. After adjusting the tracking speed of the RA drive we 
were able to keep the star image centroid within one pixel for at least 20 
minutes, much longer than expected exposure time of 3 minutes. 

The PSF size and shape were therefore completely determined by the lens 
quality, which turned out to be of moderate quality. The best focus PSF 
(Fig.~1) consisted of the sharp, asymmetric (triangular) peak overlayed on the 
extended wings containing pronounced "fingers". Although its average FWHM is 
only 2.5~pixels, bright stars' wings are easily traceable to more than 15 
pixels from the centroid. Defocusing the lens did not help since it influenced 
mainly extended feature size. 

\section{Photometry Tests at the Las Campanas Observatory}
In March 1997, after obtaining a kind approval from the Carnegie Institution 
of Washington, we have moved our instrument to the Las Campanas Observatory, 
Chile, which is operated by the Carnegie Institution of Washington. It was 
placed in the vicinity of the new telescope (Udalski, Kubiak and Szyma{\'n}ski 
1997) and room for the control computer was kindly allocated in its control 
building. 

The first testing run started on April 4, 1997 and lasted till April 26, 1997. 
During the first nights we were able to setup hardware and software, adjust 
the instrument, take calibration frames and prepare routine observations. 

For the test run over twenty  ${2\times3}$~deg fields were selected. They 
included standard calibration fields (\eg PG1323-086), crowded fields (in 
Milky Way and LMC), galaxy rich area (Virgo), ecliptic fields and star-poor 
regions (Octans). 

\subsection{Aperture {\em vs.} Profile Photometry Tests}
Two fields were used for photometry tests: crowded field in Centaurus 
(\coo{2000}{11}{35}{00}{-60}{00}{00}) ) and quite empty field close to the 
standard field PG1323-086 (\coo{2000}{13}{25}{00}{-08}{50}{00}). Over 5000 
stars were detected in the Centaurus field and only 600 in the PG1323-086 
field. 

We have tried both aperture photometry (using our own, combined program for 
detection, classification and photometry of objects) and profile photometry 
(DAOPHOT, Stetson 1987) programs. We have also tried DoPhot (Schechter, Mateo 
and Saha 1993) but without much success because of our non-symmetric PSF. 

\begin{figure}[htb]
\vspace*{8.5cm}
\includegraphics{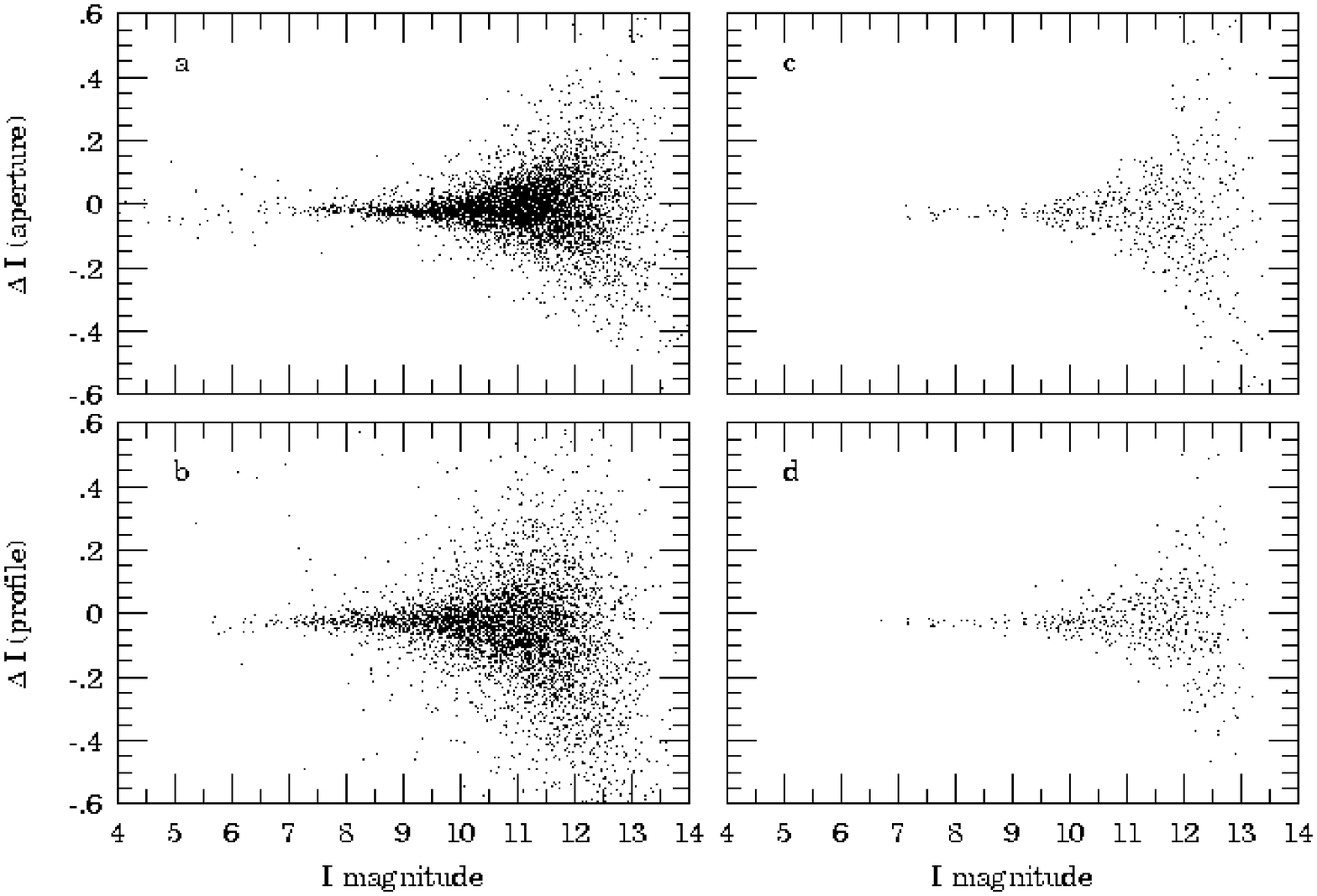}
\FigCap{Magnitude difference measured for the same stars found on two images 
of the same field: a)~and b)~aperture and profile photometry of crowded 
Centaurus field, c)~and d)~aperture and profile photometry of non-crowded 
PG1323-086 field.} 
\end{figure}

\begin{figure}[htb]
\vspace*{5cm}
\includegraphics{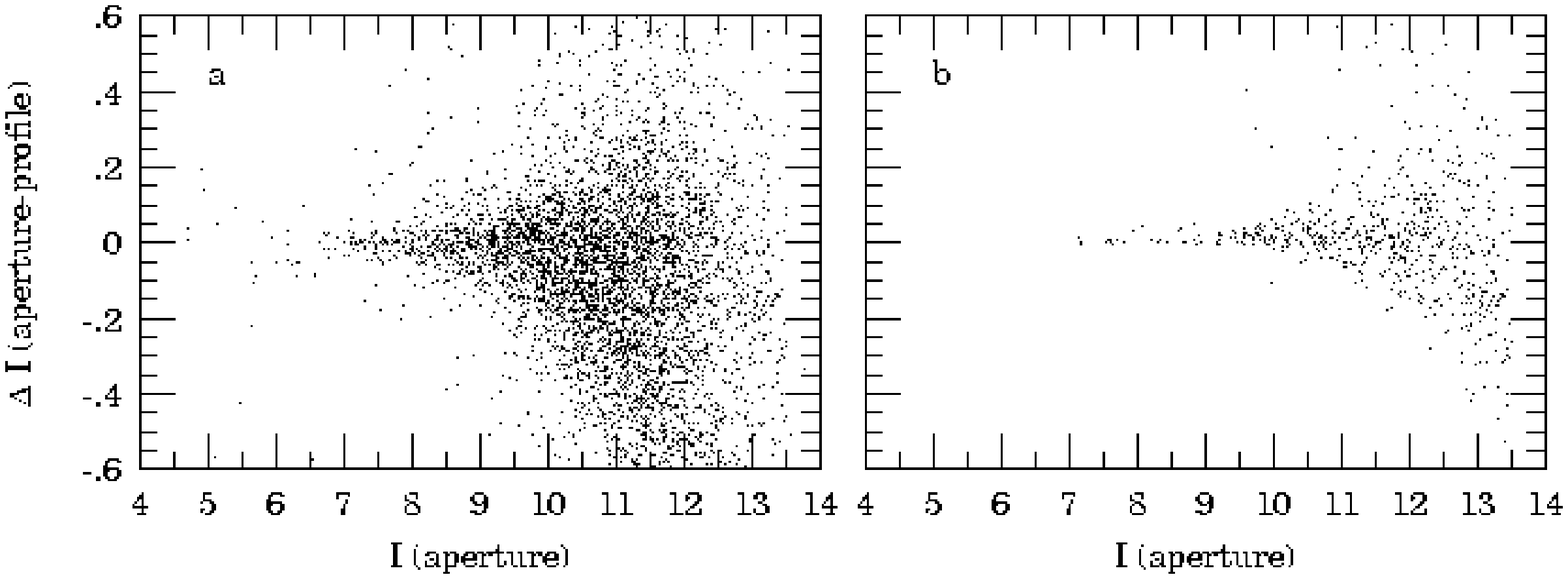}
\FigCap{Difference between profile and aperture photometry
for the same frame: a) Centaurus field, b) PG1323-086 field.}
\end{figure}

Fig.~2 presents magnitude difference {\em vs.} instrumental $I$ magnitude for 
the same stars measured on two images. Left diagrams (a,b) were obtained for 
the crowded Centaurus field, while right ones (c,d) for PG1323-086 frames. 
Upper figures (a,c) were obtained using aperture and the lower ones (b,d) 
using profile photometry. 

It is clear from Figs.~2a and 2b, that for crowded fields aperture photometry 
works much better, at least as far as measurement repeatability is concerned. 
In empty fields this tendency is reversed -- profile photometry gives slightly 
smaller magnitude dispersion. 

Fig.~3 shows difference between aperture and profile photometry for the 
crowded~(a) and empty~(b) fields. In the non-crowded field the dispersion of 
differences is of the same order as dispersion of measurements obtained using 
either of methods. There is also a clear linear correspondence between both 
results. The crowded field data exhibit much larger dispersion  and show that 
for fainter stars linear correspondence between both methods breaks. Both 
effects are caused partly by aperture photometry errors in crowded field and 
partly by profile photometry problems in the case of bad PSF. 

Since, at least at the present stage of our project, we are much more 
interested in the variability detection than in the absolute photometry, we 
decided to use more repeatable aperture photometry. This gives us also strong 
computational advantage since the aperture photometry in the crowded fields is 
by factor of~15 faster than profile fitting method. 

\begin{figure}[htb]
\vspace*{6.5cm}
\includegraphics{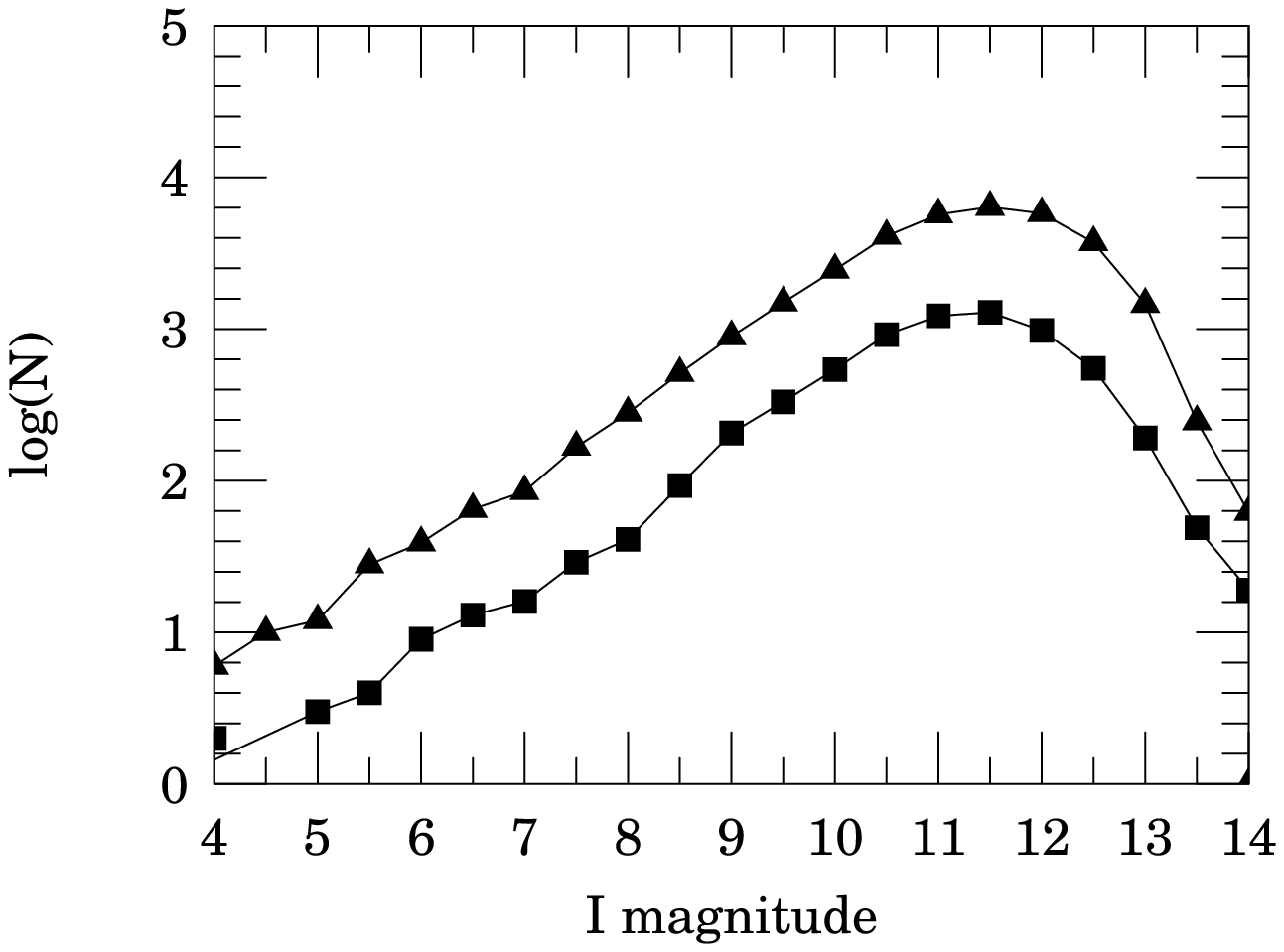}
\FigCap{Number of stars detected in 0.5~mag $I$-band bins in Centaurus 
field (squares) and on all program fields (triangles).} 
\end{figure}

Stars brighter than 7.5~mag usually contain saturated pixels. However, thanks 
to the special treatment by our aperture photometry program, we are able to 
measure also saturated stars, up to about ${I\sim3}$~mag. Obviously rms error 
of such observations rises significantly and they cannot be brought onto the 
standard system using general transformation. Nevertheless, differential 
photometry with ${\sim0.08}$~mag error is feasible. 

\MakeTable{|r|l|r|r||r|l|r|r|}{8cm}
{Standard deviation envelope of $I$-band measurements and number of stars 
detected in 0.5~mag bins in Centaurus field and in all program fields.} 
{
\hline
$I$ & $\sigma_{I}$ & $N_{\rm Cen}$ & $N_{\rm all}$ & $I$ & $\sigma_{I}$ & 
$N_{\rm Cen}$ & $N_{\rm all}$ \\
\hline
5.0& 0.07  & 3   & 12&  9.5 & 0.009 & 331  &1495\\
5.5& 0.07  & 4   & 28& 10.0 & 0.015 & 540  &2494\\
6.0& 0.07  & 9   & 39& 10.5 & 0.019 & 919  &4109\\
6.5& 0.07  & 13  & 65& 11.0 & 0.032 & 1222 &5696\\
7.0& 0.06  & 16  & 85& 11.5 & 0.045 & 1288 &6378\\
7.5& 0.008 & 29  &167& 12.0 & 0.069 & 979  &5785\\
8.0& 0.008 & 41  &280& 12.5 & 0.105 & 550  &3728\\
8.5& 0.008 & 93  &511& 13.0 & 0.176 & 191  &1459\\
9.0& 0.008 & 206 &895& 13.5 & 0.315 & 49   &246\\
\hline
}

In Fig.~4 number of stars detected in 0.5~mag $I$-band bins in Centaurus 
field and in all program fields are plotted. Only stars with more than 20 
measurements were counted. Data are also listed in Table~1. It is clear, that 
our present survey is complete to about ${I\sim11}$~mag, and includes more 
than 30\% of 12~mag stars. 

\Section{Photometry Results}
Fig.~5 shows a plot of the standard deviation in $I$-band {\em vs.}\ $I$ 
magnitude obtained for over 6000 stars with at least 50 measurements in 160 
frames of the Centaurus field. Dispersion increase for stars brighter than 
about 7~mag caused by saturation, and fast rise for stars fainter than 
13~mag are clearly visible. Numbers corresponding to the lower envelope of 
the data points are listed in Table~1. 
\begin{figure}[htb]
\vspace*{6.5cm}
\includegraphics{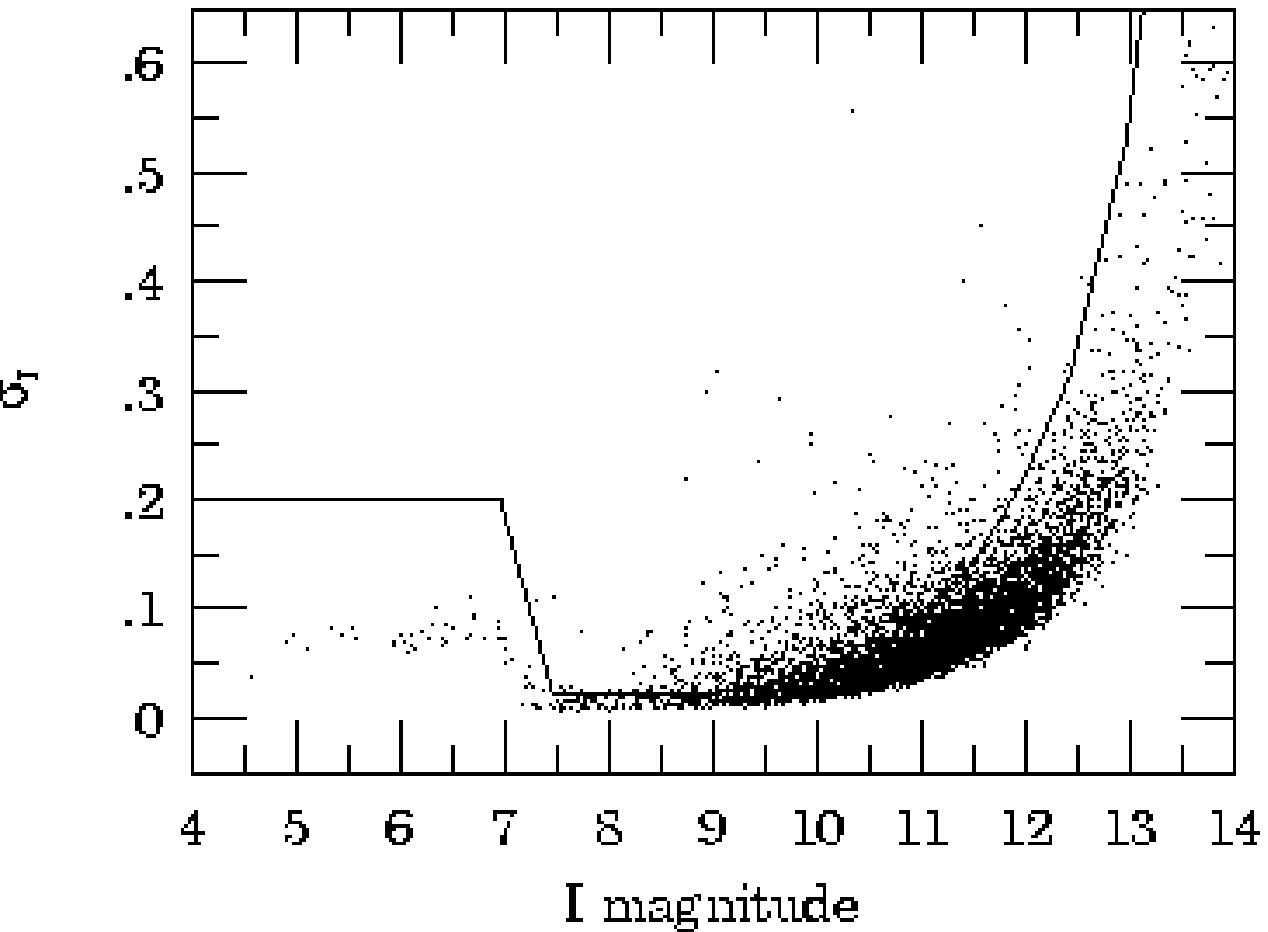}
\FigCap{Standard deviation of $I$-band magnitudes {\em vs.}\ $I$ magnitude in 
the Centaurus field. Continuous line delimiters stars which standard 
deviation is larger than three times the envelope value.} 
\end{figure}

In a search for periodic variable stars we restrict ourselves to the stars, 
that have standard deviation 2 or 3 times larger than the envelope. The number 
of stars that deviate more than that is still large in Fig.~5. This may 
indicate some problems with long-term photometric stability.

In Fig.~6 light curves of 8 stars with brightness ${I\sim}$ 6, 7 (saturated), 
8, 9, 10, 11, 12 and 13~mag are plotted. Only stars with small standard 
deviations (close to the envelope of data points in Fig.~5) were selected, and 
indeed, they look pretty constant. 

Unfortunately this is not the case for all the stars. We have found some cases 
of the small correlated variations of stellar brightness. An example of such 
situation is presented in Fig.~7. It shows 6 light curves of a set of 15 stars 
laying within 15 arc min, for which magnitudes were obtained differentially, 
so that the mean magnitude of the set remains constant. One can see, that some 
correlated (and, of course, anticorrelated) variations with amplitude about 
0.02~mag occur with the time scales of a few days. 

We have looked for possible explanations of this instability but did not find 
definite answers. The most likely reason is that the sky flat images do not 
provide proper sensitivity calibration in case of many internal reflections in 
the lens and improper baffling. We will address this problem during the next 
run. 

\begin{figure}[p]
\vspace*{10.2cm}
\includegraphics{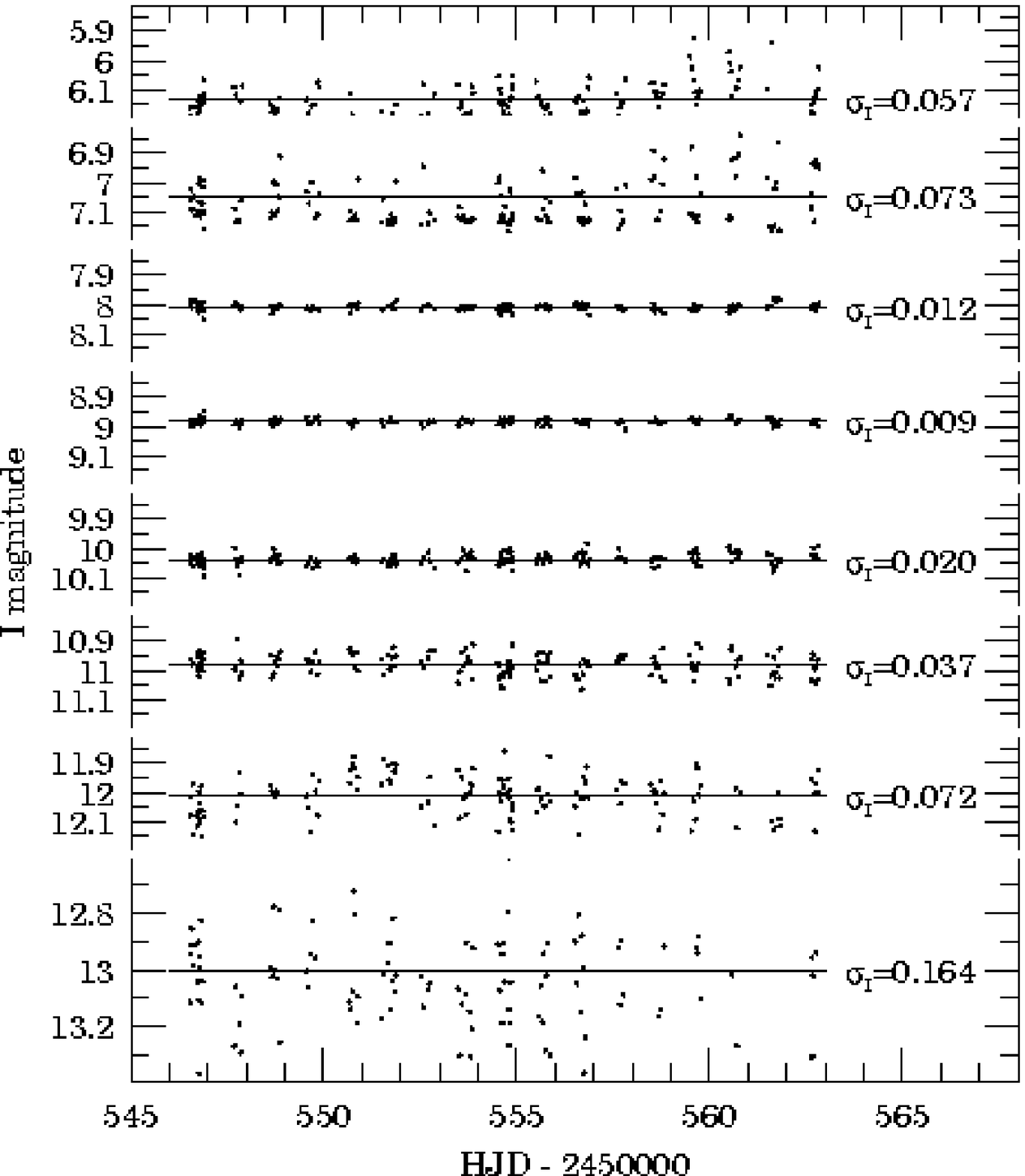}
\FigCap{Light curves for 8 non-variable stars in the magnitude range 
${I\sim6~{\rm mag}-13~{\rm mag}}$. Abscissa values cover 0.4~mag ranges for 
each star (0.8~mag for 13~mag star). Two brightest stars are saturated. Stars 
were selected to have standard deviations ($\sigma_{I}$) close to the envelope 
from the Fig.~5.} 
\vspace*{7.8cm}
\includegraphics{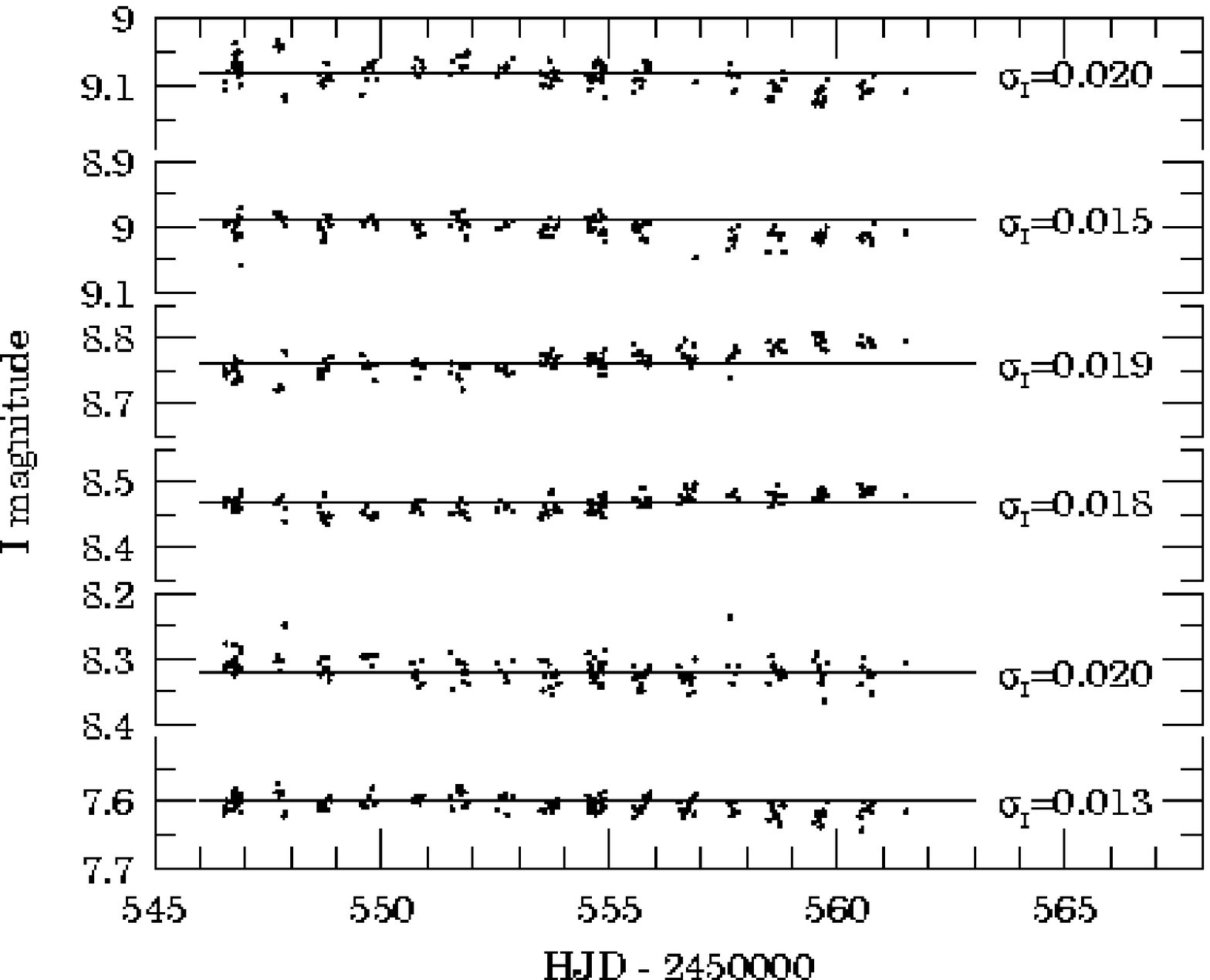}
\FigCap{Light curves of 6  out of 15 non-variable stars laying with 15 arc 
min. Raw stellar magnitudes were transformed to the common system in which 
average magnitude of the set does not change. Some long-term, correlated 
variations are present.} 
\end{figure}

\begin{figure}[htb]
\vspace*{8.5cm}
\includegraphics{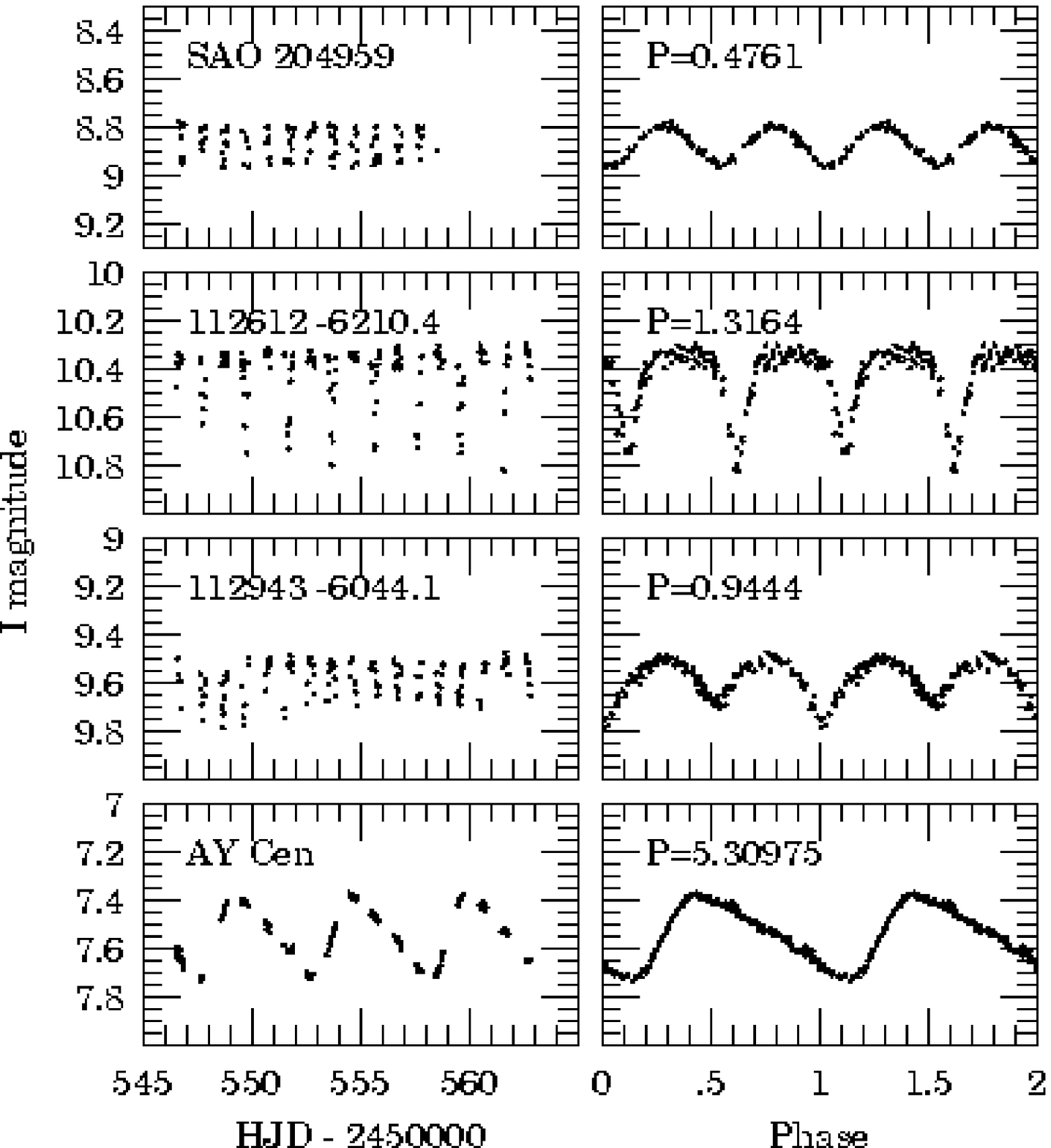}
\FigCap{Four examples of the periodic variables detected during the Las 
Campanas test run. Left diagrams show $I$-band magnitude {\em vs.} 
heliocentric Julian Data (HJD), while the right ones -- phased light curves. 
Periods were determined using PDM technique. Only AY~Cen was previously known 
to be variable.} 
\end{figure}

Very crude search through our catalog revealed over 70 variables with periods 
shorter than ${\sim10}$~days; about 50 of them are missing in the General 
Catalog of Variable Stars (Kholopov \etal 1985). Fig.~8 shows four examples of 
the periodic variables detected above 3 standard deviation threshold using 
automatic phase dispersion minimization (Stellingwerf 1978) program. Two 
cycles with zero phase chosen arbitrary are shown. The bottom curve is a 
seventh magnitude cepheid AY~Cen (\coo{}{11}{25}{06}{-60}{44}{04}, 
${P=5\zdot\upd30975}$). Three other are newly discovered quite bright 
variables: W~UMa type (\coo{}{11}{29}{43}{-63}{23}{09} ${P=0\zdot\upd9444}$), 
Algol type (\coo{}{11}{26}{12}{-62}{10}{14}, ${P=1\zdot\upd3164}$) and 
SAO~204959 (\coo{}{13}{53}{40}{-30}{36}{02}, ${P=0\zdot\upd4761}$), other 
W~UMa type star. 

Detailed catalog of the variable stars detected during the Las Campanas test 
will be published elsewhere. At present it is available over the Internet:\\
http://www.astrouw.edu.pl/$\sim$gp/asas/variables.html. 

\Section{Conclusions and Prospects}
Using a very simple and inexpensive equipment we were able to monitor large 
area of the sky in the fully automatic way. 

During our test run at the Las Campanas Observatory we have monitored over 150 
sq.\ degrees of the sky. Each night 5 to 10 exposures of each field were 
obtained, resulting in over 30~000 stars observed more than 50 times each. In 
present configuration (14~arcsec/pixels) our observations were sky limited, 
but nevertheless 12~mag limit was easily achieved using 135/1.8 telephoto 
lens. Photometry results were acceptable although some unidentified source of 
small errors excludes at the moment precise data calibration. We estimate, 
that differential photometry is accurate to about 0.02~mag for the brightest 
stars, but we cannot bring our current $I$-band magnitudes closer than 0.05 
to the standard values. 

The next step of our project will be to exchange our small (${512\times768}$) 
CCD detector for the much larger one (${\rm 2k\times2k}$), what will 
immediately result in 10 times increased data flow. We will upgrade our 
computing power to be able to reduce data flow of 1~GB per night. This way we 
will be able to monitor over 6000 sq.\ degrees per night with only one 
instrument. Once we get it working we are going to clone the system and place 
several copies of it in a few good observing sites. Having a few instruments 
running we will use different filters, to obtain, not available now, 
multiwavelength photometry. 

We are going to reduce human interaction with the system to the minimum. Next 
generation instruments will be protected by automatic, self controlled 
enclosures. Users will interact remotely only with the final photometric 
database. We also plan to add an early detection system to our database 
software to detect that exceptional events in the real time.  

\Acknow{It is a great pleasure to thank Prof.\ Bohdan Paczy{\'n}ski for 
initiative in this project, providing necessary funds and creative 
contribution at all stages of the project. 

We would like to thank Drs.\ Augustus Oemler and Miguel Roth from the 
Observatories of the Carnegie Institution of Washington for letting us use the 
Las Campanas Observatory facilities and offering invaluable support at the 
site. 

We are indebted to Drs.\ Marcin Kubiak and Andrzej Udalski for letting us use 
facilities of the telescope to house our prototype device, to Dr.\ Micha{\l} 
Szyma{\'n}ski for programming advice and computer system set up, and to all 
members of the OGLE team for continuous care of our instrumentation.

This work was partly supported by the KBN BST grant.}

\end{document}